\documentclass[prl,rd,showpacs,twocolumn,superscriptaddress]{revtex4-1}
\usepackage{amsfonts,amsmath,amssymb,amsthm}
\usepackage[bookmarks, bookmarksopen, bookmarksnumbered]{hyperref}

\usepackage{graphicx}
\usepackage{color}
\usepackage{siunitx}
\usepackage[utf8]{inputenc}
\usepackage[T1]{fontenc}
\usepackage{lmodern}
\usepackage[right]{rotating}
\usepackage{longtable}
\usepackage{array}
\usepackage{multirow}
\usepackage{paralist}
\usepackage{colortbl}

\newcommand{\codename}[1]{\texttt{#1}}

\renewcommand{\i}{\mathrm{i}}

\graphicspath{{figs/}}

\usepackage[normalem]{ulem}

\begin{document}

\title{Convective Excitation of Inertial Modes in Binary Neutron Star Mergers}
\date{\today}

\author{Roberto \surname{De Pietri}}
\affiliation{Parma University, Parco Area delle Scienze 7/A, I-43124 Parma (PR), Italy}
\affiliation{INFN gruppo collegato di Parma, Parco Area delle Scienze 7/A, I-43124 Parma (PR), Italy}

\author{Alessandra \surname{Feo}}
\affiliation{Department of Chemistry, Life Sciences and Environmental
Sustainability, Parma University, Parco Area delle Scienze, 157/A, I-43124 Parma (PR), Italy}
\affiliation{INFN gruppo collegato di Parma, Parco Area delle Scienze 7/A, I-43124 Parma (PR), Italy}

\author{Jos\'e A.~\surname{Font}}
\affiliation{Departamento de
  Astronom\'{\i}a y Astrof\'{\i}sica, Universitat de Val\`encia,
  Dr. Moliner 50, 46100, Burjassot (Val\`encia), Spain}
\affiliation{Observatori Astron\`omic, Universitat de Val\`encia, C/ Catedr\'atico 
  Jos\'e Beltr\'an 2, 46980, Paterna (Val\`encia), Spain}

\author{Frank \surname{L\"offler}}
\affiliation{Heinz-Nixdorf Chair for Distributed Information Systems, Friedrich Schiller University Jena, 07443 Jena, Germany}
\affiliation{Center for Computation \& Technology, Louisiana State University, Baton Rouge, LA 70803 USA}

\author{Francesco \surname{Maione}}
\affiliation{Parma University, Parco Area delle Scienze 7/A, I-43124 Parma (PR), Italy}
\affiliation{INFN gruppo collegato di Parma, Parco Area delle Scienze 7/A, I-43124 Parma (PR), Italy}

\author{Michele \surname{Pasquali}}
\affiliation{Parma University, Parco Area delle Scienze 7/A, I-43124 Parma (PR), Italy}
\affiliation{INFN gruppo collegato di Parma, Parco Area delle Scienze 7/A, I-43124 Parma (PR), Italy}

\author{Nikolaos \surname{Stergioulas}}
\affiliation{Department of Physics, Aristotle University of Thessaloniki, Thessaloniki 54124, Greece}

\begin{abstract}
We present the first very long-term simulations
(extending up to $\sim$\SI{140}{ms} after merger)
of binary neutron star mergers with piecewise polytropic equations of state and in full
general relativity. Our simulations reveal that at a time of 30-50 ms
after merger, parts of the star become convectively unstable, which
triggers the excitation of inertial modes. The excited inertial modes
are sustained up to several tens of milliseconds and are potentially observable
by the planned third-generation gravitational-wave detectors at frequencies
of a few kilohertz. Since inertial modes depend on the rotation rate of the
star and they are triggered by a convective instability in the
postmerger remnant, their detection in gravitational waves will
provide a unique opportunity to probe the rotational and thermal state
of the merger remnant. In addition, our findings have implications for
the long-term evolution and stability of binary neutron star remnants.
\end{abstract}

\LTcapwidth=\columnwidth


\maketitle

{\it Introduction.---}
On August 17, 2017, the Advanced LIGO and
Advanced Virgo detectors observed the first gravitational-wave (GW)
signal produced by the merger of two neutron stars in a binary system,
GW170817~\cite{Abbott:2017a}. This landmark detection initiated the
field of GW multimessenger astronomy~\cite{Abbott:2017b} in which,
thanks to the unprecedented coordinated action of LIGO, Virgo and some
70 astronomical facilities, both ground-based and in space, key
evidence to address open issues in relativistic astrophysics was
collected. GW170817 has probed the origin of short gamma-ray bursts
\cite{Monitor:2017mdv} and kilonovae and the $r$-process-mediated
nucleosynthesis of heavy elements \cite{Abbott:2017wuw}, along with
providing independent measures of cosmological parameters
\cite{Abbott:2017xzu}.

Numerical relativity simulations of binary neutron star (BNS) mergers
have shown that the outcome depends primarily on the masses of the
individual stars and on the equation of state (EOS)
(see~\cite{Duez2010,FaberRasio2012,Paschalidis:2016agf,Rezzolla:2017}
for recent reviews). Prompt collapse to a black hole happens above a
certain threshold mass, while delayed collapse or no collapse is
produced otherwise. In the latter cases, the remnant is a hypermassive
(HMNS)~\cite{Baumgarte:1999cq} or supramassive neutron star
\cite{CST92}, respectively. Simulations of their postmerger phase,
lasting up to a few tens of milliseconds, show the emission of significant
amounts of gravitational radiation at distinct frequencies of a few
kilohertz
(e.g.~\cite{Zhuge1994,PhysRevD.61.064001,Oechslin2002PhRvD..65j3005O,ShibataUryu2002,Shibata2005PhRvD..71h4021S,ShibataTani2006PhRvD..73f4027S,Kiuchi2009PhRvD..80f4037K,Stergioulas:2011gd,Bauswein:2011tp,Bauswein:2012ya,hotokezaka:2013remnant,Takami:2014tva,bauswein:2015unified,bernuzzi:2015modeling,dietrich:2015numerical,Dietrich:2016lyp,Dietrich:2016hky,Rezzolla:2016nxn,Lehner2016arXiv160300501L}),
with contributions as low as $\sim 1$ kHz, depending on the
EOS~\cite{Maione:2017aux}.

It was first shown in Ref.~\cite{Stergioulas:2011gd}, through a
mode analysis, that in many cases the postmerger emission includes,
apart from a dominant $m=2$ $f$-mode (denoted as $f_2$ or $f_{\rm
  peak}$), secondary, quasilinear combination frequencies between the
$f$-mode and the quasiradial, $m=0$ mode (sums and differences,
denoted as $f_{2-0}$ and $f_{2+0}$, respectively). These frequencies thus form an
equidistant triplet. The combination frequencies are primarily present
in soft EOS models with high mass \cite{bauswein:2015unified}. In addition,
in~\cite{bauswein:2015unified} it was found that in stiff EOS
models with low mass a different secondary peak, $f_{\rm spiral}$, due to a spiral
deformation excited during merger, can have comparable or larger
amplitude than $f_{2-0}$ and a frequency systematically larger than
$f_{2-0}$.  The detection of GWs in the postmerger remnant can lead
to tight constraints on the neutron star EOS, primarily through the
application of an empirical relation between the $m=2$ $f$-mode
frequency and the neutron star
radius~\cite{Bauswein:2011tp,Bauswein:2012ya,Clark:2014wua,bauswein:2015unified,Clark:2015zxa,BSJ,2017arXiv171100040C,Bose:2017jvk,Yang:2017xlf}
(see \cite{Abbott:2017c} for the first search of postmerger GWs in
GW170817). For additional studies of the properties of postmerger
remnants, see
e.g.~\cite{2008PhRvD..78h4033B,2008PhRvD..77b4006A,2008PhRvD..78b4012L,Kastaun:2014fna,DePietri:2015lya,Kastaun:2016yaf,PEPS2015,EPP2016,Hanauske:2016gia,Foucart:2015gaa,Maione:2017aux}.

\begin{figure*}
\includegraphics[width=0.94\textwidth]{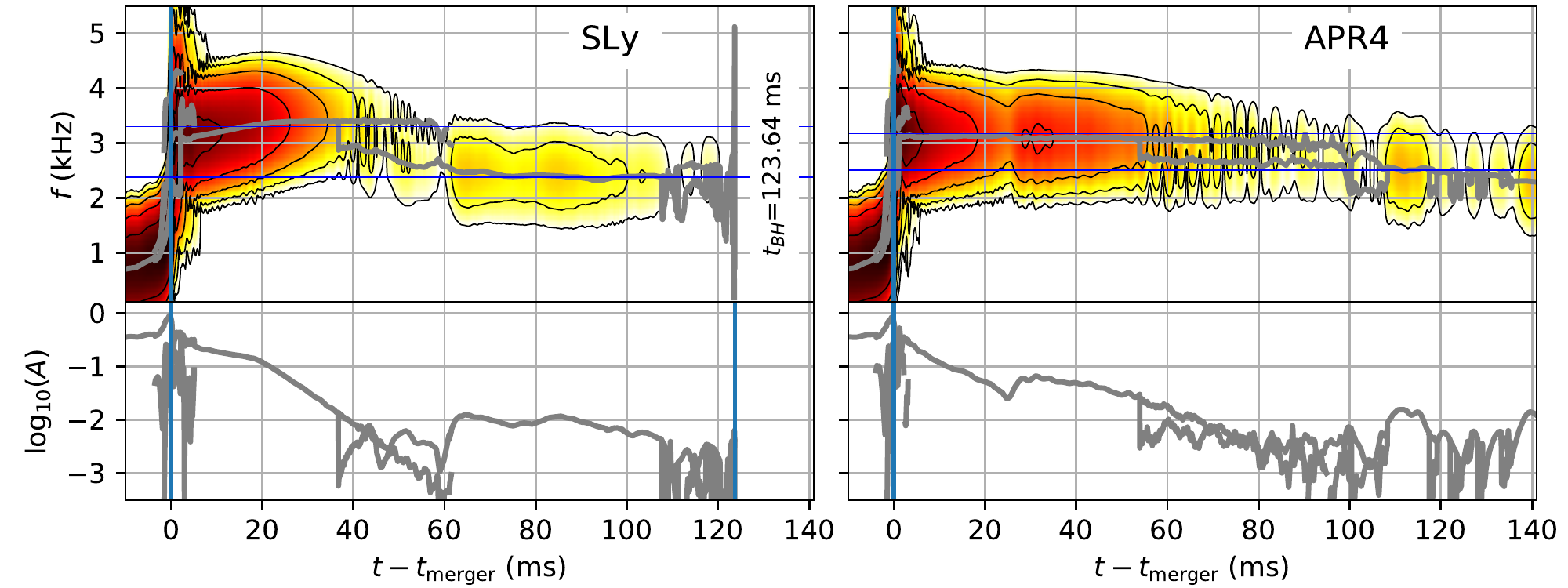}
\vspace{-0.3cm}
\caption{Top panels: time-frequency spectrograms of the $h_{22}$
  component of the GW signals for both EOS. The thick gray lines
  indicate the frequency of the active spectral mode responsible for
  the GW emission. Colors indicate the relative intensity of the
  spectral density (darker areas correspond to higher intensity).
  Bottom panels: mode amplitudes (in arbitrary units).}
\label{fig:mainSPECT}
\end{figure*}
  
In previous simulations, the dynamics of the remnant was followed up
to $\sim 40$ ms after the merger. In this {\it Letter} we present the
first very long-term simulations of BNS mergers (reaching up to $\sim
140$ ms after merger) using a piecewise polytropic EOS with a thermal
part. Our simulations reveal that at $t-t_{\rm merger}\sim 30-50$ ms
(depending on the model), the initial $m=2$ $f$-mode oscillation has
diminished and parts of the star become convectively
unstable. Subsequently, the convective instability triggers the
excitation of global, discrete inertial modes, in which the Coriolis
force is the dominant restoring force. The inertial modes are
sustained up to several tens of ms and will detectable by the planned
third-generation GW detectors at frequencies of a few kHz. We note
that these are the first simulations of BNS mergers that show the
excitation of inertial modes.

Detailed perturbative studies will be required to identify the
particular inertial modes excited in BNS mergers, which could be
either polar-led (gravito-inertial) or axial-led (similar to
$r$-modes); see~\cite{FL:1999} for definitions. Since inertial-mode
frequencies depend on the rotation rate of the star and their
existence in the postmerger phase is triggered by a convective
instability, their detection in GWs will provide a unique opportunity
to probe the rotational and thermal state of the merger remnant. Our
findings also have implications for the long-term evolution and
stability of BNS remnants.

{\it Initial data and methods.---}
Initial data were generated with the \codename{LORENE} code
\cite{Lorene:web,Gourgoulhon:2000nn}. We employ two EOS, namely the
APR4 and SLy EOS, parametrized as piecewise
polytropes~\cite{Read:2009constraints} with 7 pieces plus a thermal
component with adiabatic index $\Gamma_{\rm th} = 1.8$.  
We consider systems of total mass of 2.56~M$_{\odot}$, with an initial separation
of roughly \SI{44.3}{km} (four full orbits before merger) 
and an initial angular velocity of $\simeq$
\SI{1770}{s^{-1}}. In particular, we consider equal-mass systems characterized 
by relatively low-mass components, below the range of the inferred masses for
GW170817~\cite{Abbott:2017a} (total mass between 2.73 and 3.29
M$_{\odot}$ for conservative spin priors).
The compactness of the stars is 0.166 for the APR4
EOS and 0.161 for the SLy EOS.  The remnant HMNS survives for more
than \SI{100}{ms} before collapsing to a black hole, developing
interesting new dynamics, as we show below.
The simulations were performed using the \codename{Einstein Toolkit}~\cite{Loffler:2011ay} 
employing the same setting
of~\cite{DePietri:2015lya,Maione:2016zqz,Feo:2016cbs,Maione:2017aux}. The
only difference is the use of $\pi$-symmetry to reduce the
computational cost by a factor of 2 (but one model was also compared to a
simulation without $\pi$-symmetry). This allowed us to extend the
limit of the simulation time up to $\sim$ \SI{160}{ms}, of which the
last \SI{140} {ms} correspond to the postmerger phase.  
The reported simulations employed a finest resolution 
of $\Delta x\simeq 275$ m.  The outermost grid boundary was set at 
$\sim1040$ km from the center of mass of the system. Despite the long
evolution times of our simulations, the numerical violations of the
Hamiltonian and momentum constraints converge away at the expected 2nd
order upon increasing the grid resolution. Convergence, as well as the 
overall behavior of the evolutions and the appearance of convective instabilities
and different modes at specific times, has been checked by
performing additional simulations employing grid resolutions of $\Delta x=369$ m
and $\Delta x=185$ m (at the innermost refinement level) for selected
models. Details will be reported in~\cite{DePietri:2018}.

{\it Results overview.---}The general dynamics of our BNS mergers is
summarized in Figure~\ref{fig:mainSPECT} for both EOS. The upper
panels show time-frequency spectrograms of the $h_{22}$ component of
the spherical-harmonic decomposition of the GW signals. We superimpose
the frequency (thick gray lines) of the dominant spectral modes which
we determine using the ESPRIT Prony's method (using a moving window
interval of \SI{3}{ms}) as discussed in \cite{Maione:2017aux}. In the
lower panels, we show the time evolution of the corresponding
extracted amplitudes (in arbitrary units). For the SLy EOS, the
postmerger remnant survives for over 120~ms before collapsing to a
black hole. For the APR4 EOS, the remnant has not collapsed even
after~140 ms.

The dominant mode in the {\it early} postmerger phase is the $m=2$
$f$-mode, with a frequency above 3 kHz for both EOS and a decaying
amplitude. However, we find that a {\it distinct}, lower-frequency
mode appears later in the evolution, at $t-t_{\rm merger}\sim 35$ ms
for SLy and $\sim 55$ ms for APR4. At later times, the new mode
becomes the dominant emitter of GWs.

\begin{figure}
\includegraphics[width=0.42\textwidth]{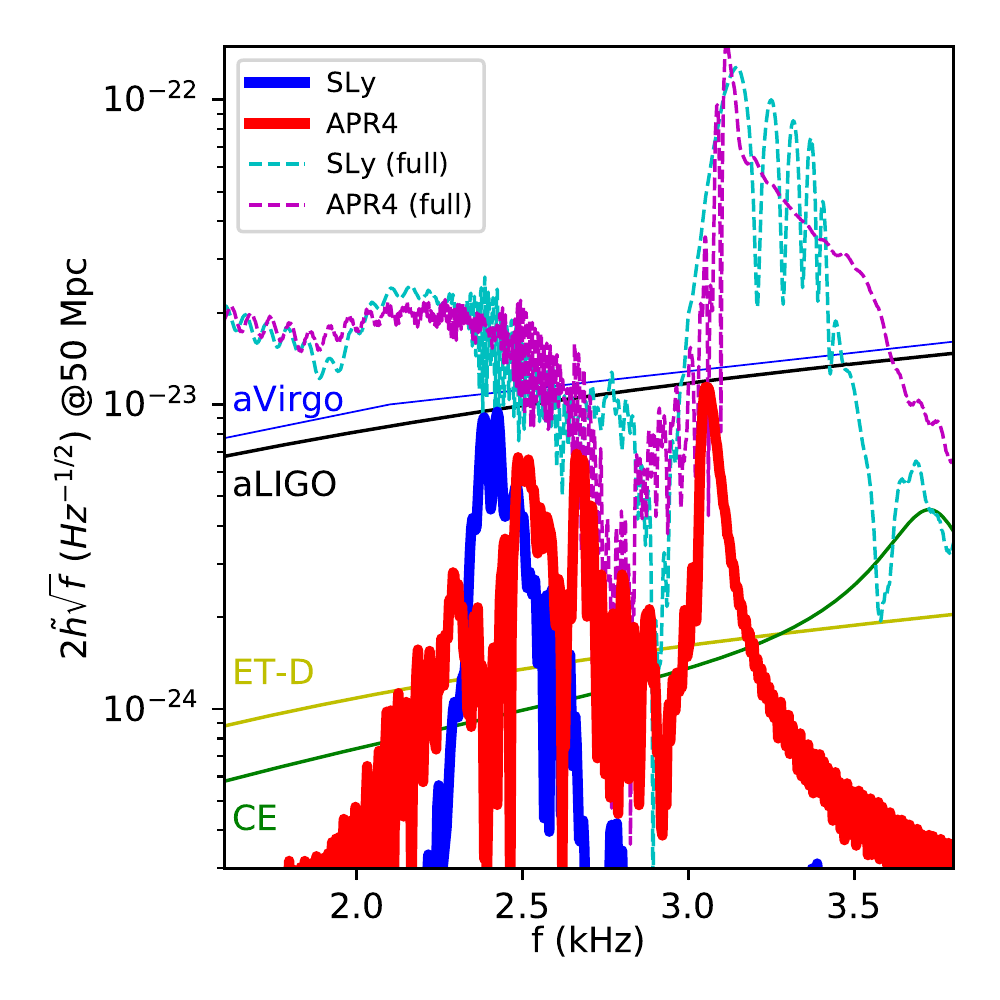}
\vspace{-0.5cm}
\caption{GW spectrum for a BNS merger at 50 Mpc at the optimal orientation. It is shown for the entire GW 
signal (dashed lines) and for a restricted time window (see main text) 
where inertial modes are active (solid lines).  
The design sensitivities of Advanced Virgo~\cite{TheVirgo:2014hva}, 
Advanced LIGO~\cite{TheLIGOScientific:2014jea}, 
Einstein Telescope~\cite{Punturo:2010zz}, and 
Cosmic Explorer~\cite{Evans:2016mbw} are shown for reference.}
\label{fig:fft}
\end{figure}

\begin{figure*}
\includegraphics[width=0.90\textwidth]{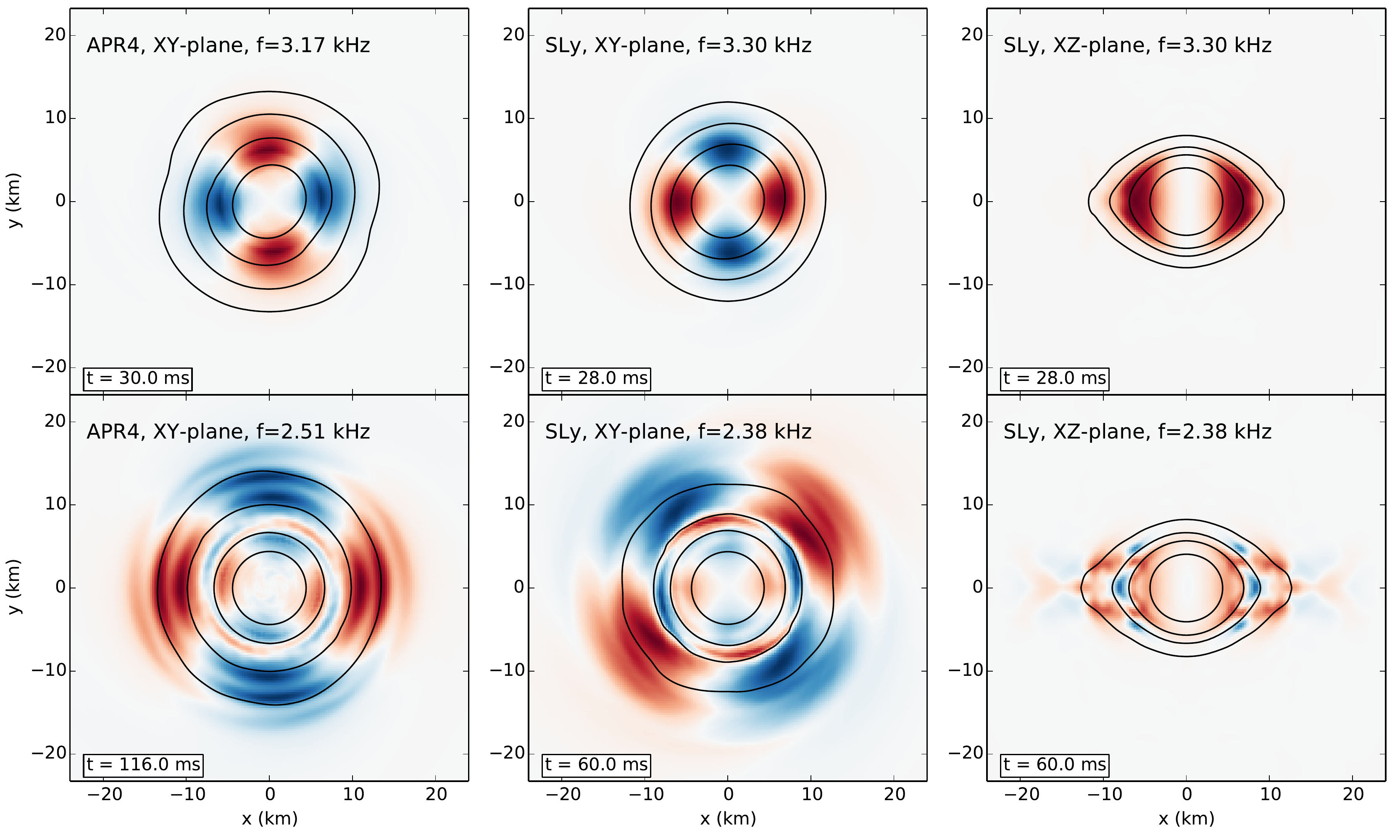}
\vspace{-0.45cm}
\caption{Left and center panels: density eigenfunctions in the
  equatorial plane for particular modes at different times (relative
  to the merger time), for the two models considered. Right panel: the
  corresponding eigenfunctions in the vertical plane for the SLy
  model. The black lines are isocontours of the rest-mass density with
  values (for decreasing radii) $1.5\times 10^{13},8\times
  10^{13},3\times10^{14}$, and $8\times 10^{14}$ (g/cm${}^3$).}
\label{fig:mainWF}
\end{figure*}

The mode dynamics is imprinted in the GW signal, whose spectrum is
displayed in Fig.~\ref{fig:fft}. We show the PSD of the GW signal for
a maximally-aligned source at a distance of \SI{50}{Mpc}, both for the
whole simulation (dashed lines), and restricted to the interval
between \SI{55}{ms} and \SI{140}{ms} after the merger (solid lines). In
the former case, GW emission dominates the spectrum between
\SI{3}{kHz} and \SI{3.5}{kHz} for the SLy EOS, and shows a single
higher peak at \SI{3.17}{kHz} for the APR4 EOS.  On the other hand,
when selecting the signal in the range between \SI{55}{ms} and
\SI{140}{ms}, the spectrum is dominated by a peak at frequency
\SI{2.38}{kHz} in the case of the SLy EOS while we see different peaks
at frequencies pf $\sim$\SI{2.3}{kHz}, $\sim$\SI{2.5}{kHz}
$\sim$\SI{2.7}{kHz} for the APR4 EOS, corresponding to inertial modes
excited at different times.
We note that there is sufficient power in these lower-frequency modes
to render them potentially observable by third-generation
detectors. For a source at 50 Mpc the expected Cosmic Explorer
\cite{Evans:2016mbw} S/N ratio for optimal use of matched filtering
techniques for the signal emitted between \SI{55}{ms} up to \SI{140}{ms} after
merger are $\sim$3.5 and $\sim$3.7 for SLy and APR4 EOS, respectively
(for the Einstein Telescope-D \cite{Punturo:2010zz} they are $\sim$2.5 and $\sim$2.9
but enhancement can be expected due
to the triangular arrangement with three non-aligned
interferometers, see \cite{Yang:2017xlf}).
A first insight into the nature of this late-time feature can be
obtained by examining the eigenfunction of the mode, which we extract
by taking the FFT of the time series of the rest-mass density in the
equatorial plane and plotting the amplitude at a specific frequency
(see~\cite{SAF:2004}). The first row of Figure~\ref{fig:mainWF} shows
the eigenfunction of the dominant $m=2$ $f$-mode at an early time, for
the two models. These eigenfunctions have no azimuthal nodal line,
corresponding to a fundamental mode. The last panel in the upper row
of Figure~\ref{fig:mainWF} shows the extracted eigenfunction for the
SLy model in the vertical plane, which is consistent with an $m=2$
$f$-mode \cite{2013FSbook}.

The second row of Figure~\ref{fig:mainWF} shows the eigenfunction of
the distinct, lower-frequency modes at a late time, for the two
models. The eigenfunctions still have an $m=2$ character in the
equatorial plane, but there are also azimuthal nodal lines, which is a
characteristic of modes with higher radial order. In addition, the
amplitude vanishes in the core and peaks at the outer layers of the
remnant. The last panel in the lower row in Fig.~\ref{fig:mainWF}
shows the extracted eigenfunction for the SLy model in the vertical
plane, which is consistent with an inertial mode~\cite{Kastaun:2008}.

\begin{figure}
\includegraphics[width=0.41\textwidth]{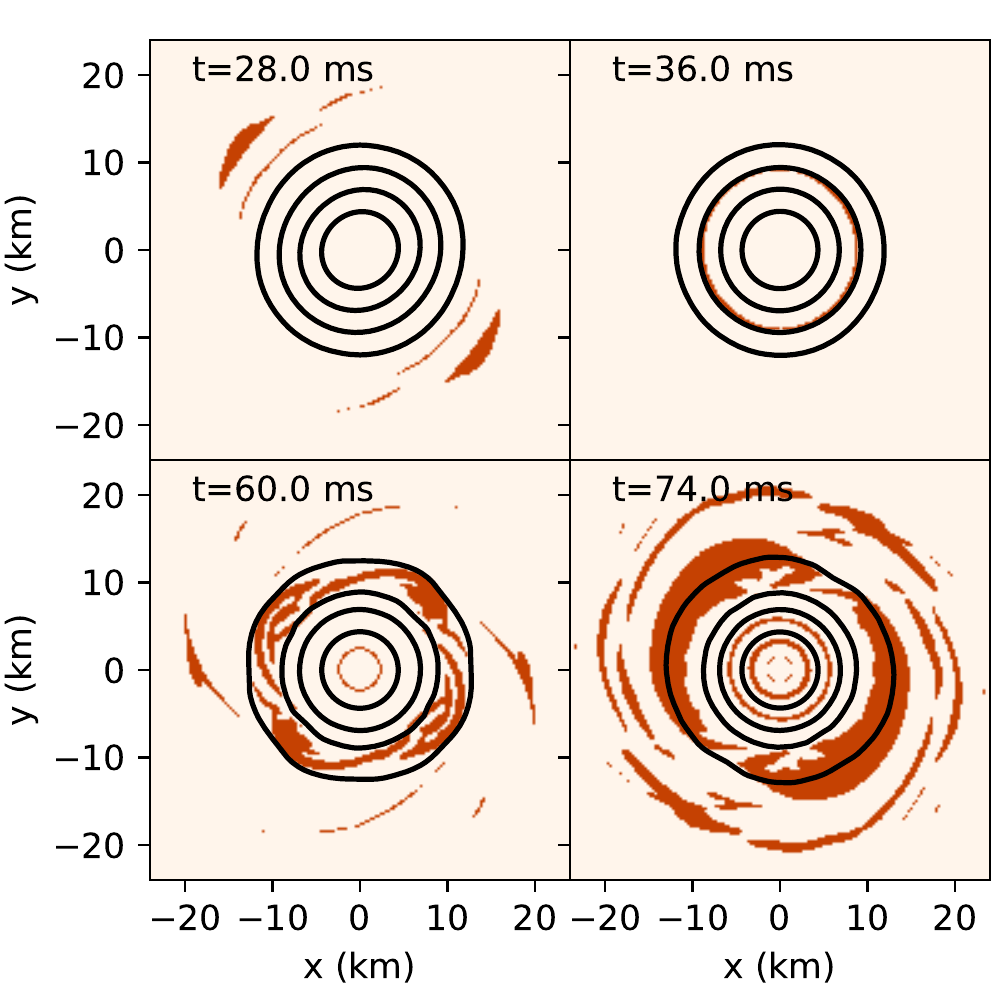}
\vspace{-0.5cm}
\caption{Snapshots of $A_r$ in the equatorial plane for the SLy model at different times relative to the merger time. Convectively unstable regions are shown with dark color.}
\label{fig.EffectiveINDEX}
\end{figure}

{\it Convective stability and inertial modes.---}The late-time,
lower-frequency modes found in our simulations can be interpreted as
{\rm inertial} modes, for which the Coriolis force is the dominant
restoring force. The growth of the inertial modes (up to their
saturation amplitude) is triggered by a convective instability that
appears in the nonisentropic remnant shortly before the inertial
modes start to grow from a small amplitude.

The local convective instability depends on the sign of the Schwarzschild discriminant
\begin{equation}
A_\alpha = \frac{1}{\varepsilon+p}\nabla_\alpha \varepsilon - \frac{1}{\Gamma_1 p} \nabla_\alpha p, 
\end{equation}
where $\Gamma_1 := (\varepsilon +p)/p (d p / d \varepsilon)_s = (d \ln p / d \ln \rho)_s$ is the adiabatic index of linear perturbations about a pseudobarotropic equilibrium (see \cite{2013FSbook} and references therein) and where $\varepsilon$ is the energy density. Regions with $A_\alpha<0$ are convectively stable, whereas regions with $A_\alpha>0$ are convectively {\it unstable}. 
For a piecewise polytropic EOS we calculate $\Gamma_1$ as
\begin{eqnarray}
 \Gamma_1 
 &=& \Gamma_{\rm th}+(\Gamma_i-\Gamma_{\rm th})\frac{K_i\rho^{\Gamma_i}}{p},
\end{eqnarray}
where $K_i$ and $\Gamma_i$ are the polytropic constant and exponent in the $i$-th 
piece of the EOS, respectively, and $\Gamma_{\rm th}$ is the adiabatic index of the added thermal part.
To determine the convective stability in the star,
we evaluate $A_r$ and $A_\theta$ in the equatorial and vertical
planes, at different times during the evolution.

Figure~\ref{fig.EffectiveINDEX} shows the convectively unstable regions,
where $A_r>0$ (dark color), in the equatorial plane for the SLy
model. The top left panel shows an instant at an early time ($t=28$
ms), when most of the star (except for some parts in the low-density
envelope) is convectively stable and the $m=2$ $f$-mode dominates the
GW spectrum. The entropy in the remnant created by shock heating 
during merger (not shown here) is concentrated in an expanding envelope 
around a relatively cold core (two initial hot spots mix into a nearly axisymmetric 
structure in the early postmerger phase, as also found by~\cite{Kastaun:2016yaf}).

At $t=36$ ms, a convectively unstable ring appears in the equatorial
plane, at $ \rho \simeq 8\times 10^{13} {\rm g/cm^3}$. This coincides
with the first appearance of inertial modes in
Fig.~\ref{fig:mainSPECT}. In accordance with the observed growth of 
convective motions, at $t=60$ ms, the first convectively
unstable ring has expanded to lower densities and appears fragmented,
which coincides with a strong growth of an inertial mode with
frequency of 2.38 kHz in Fig.~\ref{fig:mainSPECT}. The amplitude of
the inertial mode only diminishes slightly for the next 15~ms. At
$t=74$ ms, the convectively unstable regions have expanded further and
we observe a renewed growth of the amplitude of the inertial mode in
Fig.~\ref{fig:mainSPECT}. We thus observe a strong correlation between
the appearance of several convectively unstable regions in the remnant
and the appearance and growth of inertial modes.

Inertial modes can be either polar-led (gravito-inertial modes) or
axial-led ($r$-mode like). In slowly-rotating stars, gravito-inertial
modes become $g$-modes and we estimate a Brunt-V\"aiss\"al\"a
frequency $\langle N \rangle/(2\pi) = 356 {\rm Hz}$ using $\Gamma =
1.36, \Gamma_1 = 1.74$ and $\rho=8\times10^{13} {\rm g/cm^3}$
(see~\cite{2009Passamonti}). These modes are also excited in
protoneutron stars, in the context of core-collapse supernovae (see
e.g.~\cite{alex} and references therein). However, in rapidly rotating
stars, gravito-inertial modes have very similar properties to
axial-led inertial modes, with frequencies being nearly proportional
to the rotational frequency of the
star~\cite{2001Lockitch,2009Passamonti,2009Gaertig}.
The $m=2$ inertial modes identified in the simulations of the two
models discussed here have frequencies only somewhat smaller than
twice the maximum angular frequency in the differentially rotating
star $\Omega_{\rm max}$ whose rotational profile is consistent with
results found in \cite{Hanauske:2016gia,Ciolfi:2017uak} that will be
reported in \cite{DePietri:2018} (notice that the $m=2$ $f$-mode
frequency appearing at early times has been empirically associated
with $2\Omega_{\rm max}$ \cite{Kastaun:2014fna}).
The characterization of the (polar- or axial-led) inertial modes that
are excited in our simulations can be achieved after a comparison with
detailed perturbative studies, which we plan to undertake in the
future. 
We finally note that mergers with unequal mass components may
also show the excitation of odd modes, which are suppressed in our
present equal-mass simulations with $\pi$-symmetry.  For the
evolution of models very close to the ones discussed here, 
neutrino effects have been considered in~\cite{Foucart:2015gaa},
while the timescales for magnetic effects and the MRI-induced effective
viscosity have been analyzed in detail in~\cite{Kiuchi:2017zzg}.
These effects should be reconsidered on long timescales of up to 140 ms.

{\it Conclusions.---}The appearance of convectively unstable regions
and the excitation of particular inertial modes in long-lived remnants
of BNS mergers depends on the rotational and thermal state of the
remnant and impacts its dynamical evolution. 
Whereas, in this first investigation, we considered a piecewise
polytropic EOS with a thermal component (that has the advantage of having
a well-defined, analytically expressed thermodynamics) and we ignored 
the dependence on composition gradients, magnetic fields 
and neutrino transport (which should be clearly investigated in future 
simulations),  our results imply that either the presence or the 
absence of inertial modes in the late-time GW spectrum of 
BNS mergers will be a sensitive probe of neutron star physics.

\bigskip

We are grateful to Nils Andersson, Andreas Bauswein, James Alexander
Clark, Roland Haas and John Friedman for comments on the manuscript.  We thank the
many developers of the public software used for this research, namely
LORENE and the Einstein Toolkit.  Computational resources were
provided by PRACE Grant No.~Pra14\_3593, by the CINECA-INFN agreement,
by the Louisiana Optical Network Initiative (QB2, allocations
loni\_hyrel, loni\_numrel, and loni\_cactus), by the LSU HPC
facilities (SuperMic, allocation hpc\_hyrel) and by the GWAVES
(pr002022) allocation on the ARIS facility of GRNET in Athens. JAF is
supported by the Spanish MINECO (AYA2015-66899-C2-1-P), by the
Generalitat Valenciana (PROMETEOII-2014-069), and by the
H2020-MSCA-RISE-2017 Grant No.~FunFiCO-777740.  FL was supported by
the NSF in the USA as part of the Einstein Toolkit (Grants
No.~1550551, No.~1550461, No.~1550436, No.~1550514). Support by the
COST Actions MP1304 (NewCompStar), CA16104 (GWVerse) and CA16214
(PHAROS) is kindly acknowledged.

\end{document}